# REIMPLEMENTING THE EPICS STATIC DATABASE ACCESS LIBRARY*

A.N. Johnson and M.R. Kraimer, Advanced Photon Source,
Argonne National Laboratory, 9700 S. Cass Avenue, Argonne, IL 60439, USA

*Abstract*

The Static Database Access library was first introduced in EPICS[1] (Experimental Physics and Industrial Control System) Release 3.11 in 1994. It provides an application program interface (API) for database configuration tools written in C to manipulate database record definitions and is used for various tasks within the core EPICS software. This paper describes the structure of a replacement for the original library that adds a native C++ API and will make some future enhancements to EPICS significantly easier to implement.

## 1 INTRODUCTION

The Static Database Access Library (dbStaticLib) [1] provides standard facilities for EPICS and its database configuration tools (DCTs) to read and write database (db) and database definition (dbd) files, and allows the format of these files to be changed without having to modify the configuration tools each time.

Implementing Link Support [2] in EPICS base requires a number of additions to the dbStaticLib code and data structures, including changes to the existing API. This was one of the main reasons for the rewrite.

The code currently written and tested maintains database definitions in memory at design-time, reads and writes dbd files, and generates C header files from menu record and link type definitions. A new approach that will avoid having to load these definitions from the dbd file at runtime has also been coded.

## 2 LIBRARY STRUCTURE

### 2.1 Partitioning dbStaticLib

Although their APIs are essentially identical, there are actually two slightly different libraries making up the old dbStaticLib implementation. IOCs load one that stores information in a layout suitable for run-time access by iocCore and the record and device support routines. DCTs are linked against a version that holds all database field values as strings rather than as the native types needed for runtime operation.

The distinction between run-time and design time libraries might be widened after measurements of the performance and code size of the design-time C++ sections have been made. The IOC run-time code needs to be reasonably compact and perform certain search operations quickly, whereas at design time both the speed and size requirements can be relaxed in favour of flexibility. Maintaining two independent codebases would increase the software maintenance load however, so substantial commonality is desirable.

### 2.2 Object Containers

Examination of the existing source code revealed that a significant proportion consisted of housekeeping routines for creating, deleting, searching for and scanning through collections of objects of various types. Although the EPICS linked list library is used to maintain the object collections, this is only a fraction of the need, and no other code is shared between the different object types. Maintenance of this aspect of the library should be significantly reduced by using the container template facilities of the C++ standard library.

### 2.3 Parsing Files

The ASCII file parsing routines in dbStaticLib were created using the software tools flex and antelope, open source versions of the standard UNIX tools lex and yacc that are distributed with EPICS base (standard UNIX versions don't output code suitable for vxWorks due to the single symbol table that is used for all installed programs). The relatively simple and consistent syntax for db and dbd files however makes it fairly easy to write a parser for these files by hand.

The new library is designed to allow replacement of the file I/O routines that parse and output db and dbd files. In the old implementation some of the parsing routines were the only means of creating new objects of certain structured data types, making it harder to add support for other file formats such as XML. The file I/O routines in the new library are not part of the C++ class definitions, thus extra or alternative I/O routines can easily be included.

* Work supported by U.S. Department of Energy, Office of Basic Energy Sciences under Contract No. W-31-109-ENG-38.

[1] http://www.aps.anl.gov/epics

# 3 DESIGN TIME DATA

The information from a dbd file is stored in a series of C++ class objects, using standard C++ library container templates and providing public member functions to give access to the data.

The C++ compiler enforces access rules to prevent objects from being modified after being loaded. Most accessor member functions are const and/or return references or pointers to const objects. The iterators described below are all const_iterator types.

## 3.1 dbDefinition

An instance of this class holds a complete dbd file in memory and provides the ability to iterate through its contents as well as find objects by name, where appropriate. The class data members are defined roughly as follows (member functions omitted):

```
class dbDefinition {
  vector<string> drvSups;
  map<string,dbChoice*> menus;
  map<string,dbBrkTable*> brkTables;
  map<string,dbRecordDefinition*> rtDefs;
  dbLinkGroup *linkgroups[LINK_DIRECS];
  … };
```

The dbd file entry for a driver support consists of just the symbol name of the support entry table, thus the storage needs are simple. Member functions provide indexed access to these strings.

Indexed access is provided to the input, output, and forward link groups that are part of the link support code being added for release 3.15. Access to other data members is needed using both iterators and fast lookup by name, thus a standard map is used to store these.

## 3.2 Enumerations, Menus, and Breakpoints

C and C++ provide enumeration data types, but not the ability to convert between a string or character representation and its numeric value. As this is a common requirement in dbStaticLib, these operations have been encapsulated in a class template.

Several different kinds of choice menu derived from the abstract base class dbChoice provide functions to convert between string and integer representations and to return the number of different choices.

Enumerations and menus differ in that a menu object provides string conversions without keeping any state itself, while an enumeration class instance has a state. A template is provided to create a choice menu from an enumeration template instance. Choice menus are usually created from menu definitions in a dbd file though, and record DTYP fields also use a kind of choice menu described below.

The old C API for breakpoint table data becomes a C++ class dbBrkTable. A standard vector container to hold an unknown number of breakpoint intervals simplifies the code and improves the encapsulation.

## 3.3 Record Type Definitions

The main contents of a record type definition are collections of objects containing information about individual record fields and the devices and link types that support this record type. The member data are defined:

```
class dbRecordDefinition {
  const string name;
  int designFields;
  int linkFields;
  vector<dbFieldDefinition*> fields;
  map<string,int> index;
  dbDeviceMenu devices;
  dbLinkGroup linkgroup;
  … };
```

The designFields and linkFields counters are provided for convenience. The fields vector contains the field definitions, retaining the order in which they were defined and giving fast indexed access and iterator support. The index map speeds up finding field definitions by name, but may be unnecessary at design time.

A dbDeviceMenu is a choice menu that also stores the old device support entry table (DSET) symbol name and link type, as well as a choice string for users. The DTYP field will be linked to the devices menu to retain the old-style device support functionality.

Each record type also has its own group of link support definitions that are used for the 'controlling' link field for this record type, usually INP or OUT. New-style device supports will be added to this group rather than the devices menu.

## 3.4 Link Groups

A link group is a collection of link type definitions, providing iterators and by-name access to its contents.

```
class dbLinkGroup {
  map<string,dbLinkDefinition*> linkDefs;
  … };
```

Each group is associated with either a record type or one of the link "directions" DBF_INLINK, OUTLINK, or FWDLINK.

## 3.5 Link Type Definitions

A link type definition describes the data members of a C structure and is therefore very similar in concept to a record type definition. The main differences are that there are fewer possible data types (so a different field descriptor class is used), and there is no need for fast lookups of the field definitions by name. The class data members are thus minimal:

```
class dbLinkDefinition {
  const string symbol;
  const string choice;
```

```
vector<dbMember*> fields;
… };
```

The link support entry table (LSET) symbol name and a description string for users are kept with an ordered list of the fields that make up the structure. Field definitions can be accessed by index number or (using a linear search) by name.

*3.6 Describing Field Data*

A record field has more attributes than a link field, which naturally leads to the use of inheritance to define the former descriptor in terms of the latter.

```
class dbMember {
  const string name;
  const dbValType dbfType;
  dbValInfo valInfo;
  string prompt;
  string initial;
  … };
```

The dbMember class contains strings for the field name, user prompt and default value. An enumeration holds the field data type and a pointer to a dbValInfo derived object that stores additional information about the field, which is data-type specific such as number base, string length, menu type, etc.

dbValInfo is an abstract base class; a concrete derived class exists for each field type, and dbValInfo has a factory member function for creating these. The class provides string conversion and range checking for its data type, as well as storage and accessor functions for any type-specific metadata it needs.

```
class dbFieldDefinition: public dbMember{
  dbGroup guigroup;
  dbSpecial special;
  bool processPassive;
  short interest;
  dbSecurity security;
  … };
```

The dbFieldDefinition class extends dbMember, adding information about how DCTs should group fields together, special field processing, a passive processing indication for Channel Access, interest level for dbpr, and the field's access security level.

## 4 RUN-TIME DEFINITIONS

At compile time, the field definitions in a record's dbd entry are converted into a C header file containing a definition of the record structure. Also present in the old output are some conditionally compiled arrays containing field size and offset data for every field.

In the new version it is proposed to extend these arrays into data structures containing the complete record description, significantly reducing the size of the dbd file that needs to be loaded at runtime. This also stops IOCs from loading a record definition that has different fields from those expected by the record support code.

Similar static data structures and arrays are proposed for link support structure and choice menu definitions. New versions of the programs dbToRecordtypeH and dbToMenuH use the library to read a dbd file and include these static data structures in their output. A new dbToLinktypeH program similarly converts link support definitions.

Taking this idea to its extreme, a whole dbd file could be converted into C code that is compiled and linked into the IOC executable. The result would remove all code for parsing dbd files from iocCore. By using registration functions EPICS could still be extended to allow online addition of new menus, record, device, and link types.

## 5 FURTHER WORK

All the implications of compiling dbd data have not yet been developed in detail, but this approach is promising for various reasons already discussed and will be explored more fully as work on the library continues.

There is currently no code for reading or writing database files, although the requirements for doing this were considered when designing the existing code.

A wrapper also needs to be created that calls the underlying C++ code and implements as much of the old C API as possible. Some changes to existing DCTs will be needed to add support for the new link types, but we believe that the old functionality can be emulated completely using the new library.